# FIFTEEN YEARS OF ECONOPHYSICS RESEARCH

**Bikas K. Chakrabarti**

*Theoretical Condensed Matter Physics Division and*
*Centre for Applied Mathematics and Computational Science,*
*Saha Institute of Nuclear Physics, 1/AF, Bidhannagar, Kolkata 700 064, India*

*Economic Research Unit, Indian Statistical Institute,*
*203 Barrackpore Trunk Road, Kolkata 700 108, India*

Email: bikask.chakrabarti@saha.ac.in

**Anirban Chakraborti**

*Laboratoire de Mathématiques Appliquées aux Systèmes*
*Ecole Centrale Paris, 92290 Châtenay-Malabry, France*

Email: anirban.chakraborti@ecp.fr

**Abstract:** Econophysics is a new research field, which makes an attempt to bring economics in the fold of natural sciences or specifically attempts for a “physics of economics”. The term Econophysics was formally born in Kolkata in 1995. The entry on Econophysics in The New Palgrave Dictionary of Economics, 2nd Ed., Vol 2, Macmillan, NY (2008), pp 729-732, begins with “... the term ‘econophysics’ was neologized in 1995 at the second Statphys-Kolkata conference in Kolkata (formerly Calcutta), India ...”. The Econophysics research therefore formally completes fifteen years of research by the end of this year!

The importance and proliferation of the interdisciplinary research of Econophysics is highlighted in the special issue of Science & Culture, which presents a collection of twenty nine papers (giving country wise perspectives, reviews of the recent developments and original research communications), written by more than forty renowned experts in physics, mathematics or economics, from all over the world.

We present here the list of contents and the editorial. The manuscript files are available at http://fiquant.mas.ecp.fr/chakraboa for preview. This special issue will be published online at http://www.scienceandculture-isna.org/journal.htm, at the end of October 2010.

Contd. next page.

**Cover :** Trading as a scattering process

## EDITORIAL

# FIFTEEN YEARS OF ECONOPHYSICS RESEARCH

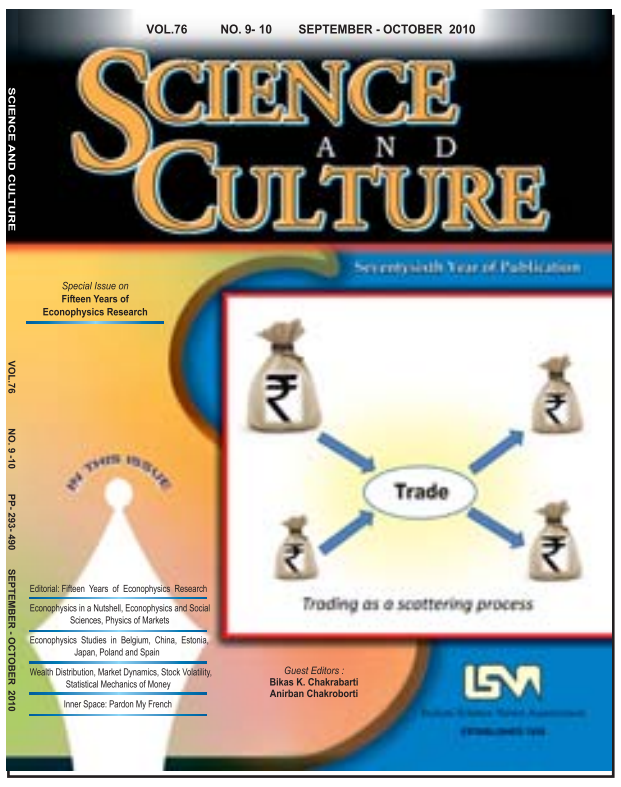

More than seventy five years back, Prof. Prasanta Chandra Mahalanobis initiated a new and pioneering experiment in India, namely founding the Indian Statistical Institute (ISI), where essentially statisticians, mathematicians and social scientists (including economists) would work to provide the country with data and models on social dynamics. It has been a premier institution of the country, contributing significantly to research and teaching of mathematical and social sciences.

> "Indian Statistical Institute (ISI) engages in the research, teaching, and application of statistics to the natural sciences and social sciences. Founded by Professor P.C. Mahalanobis in Kolkata in 1931, while statistics was a relatively new scientific field, the institute gained the status of an Institution of National Importance by an act of the Indian Parliament in 1959. ... ISI is generally regarded as being the best Indian school in the few areas of its expertise namely statistics, mathematics, computer science, quantitative economics, operations research and information science and is considered to be one of the few research oriented Indian schools at both, the undergraduate and graduate level."
>
> — Wikipedia (June, 2010)

It was truly a remarkable experiment with resounding success! In those days, both degrees awarded and subjects taught in the social sciences used to be offered by the Arts faculty, in most of the Universities (both Indian and foreign). Social sciences, including economics, was mainly considered until recently, to be part of Fine Arts, and not Science (as mathematics, statistics or computer science). As such, the ISI had been truly original, not a copy of any European or American model institution (contrary to many other institutions in India, established just before and after the independence).

This pioneering and esteemed Institute of the country, since its foundation, has remained however neglected by the various science funding agencies like the Department of Science & Technology (DST — funding basic research institutions in natural sciences), Council of the Scientific & Industrial Research (CSIR — funding applied research institutions in natural sciences), or the Department of Atomic Energy (DAE — funding major physical, mathematical, biological basic research and teaching institutions of the country), and has been funded essentially by the Planning Commission of India. Many distinguished academicians believe, it is not the most effective or desirable situation, as the Planning Commission does not support any other such basic research and teaching Institutions of the country and does not even possess the appropriate expertise or structure to do so!

In the global scenario of interdisciplinary scientific research, dramatic changes started taking place since mid-eighties. Several distinguished researchers in natural sciences and economics, started the Santa Fe Institute in USA in 1984, where several Nobel laureates in physics, biology and economics soon joined as adjunct faculty.

> "The Santa Fe Institute is a non-profit research institute located in Santa Fe (New Mexico, United States) and dedicated to the study of complex systems. It was founded in 1984 by George Cowan, David Pines, Stirling Colgate, Murray Gell-Mann, Nick Metropolis, Herb Anderson, Peter A. Carruthers, and Richard Slansky. ...Current research at the Institute focuses on a set of areas commonly described as complex systems, including both physical, evolutionary, and social systems."
>
> — Wikipedia (June, 2010)

Around this time, many universities started giving economics degrees in science and even older universities like those of Calcutta, Delhi were no exceptions! Soon, many pioneering academies of the world started electing distinguished social scientists and economists (along with natural and biological scientists) as their fellows or associates: For example, the Cambridge economist Prof. Partha Dasgupta was elected as Foreign Associate of Royal Swedish Academy of Sciences in 1991, of the US National Academy of Sciences in 2001 and Fellow of the Royal Society in 2004. This is just one example, and there are so many examples of other Indian economists becoming Fellows of distinguished science Academies of other countries. This expertise immediately helped these Academies and the agencies of their respective countries to design, take appropriate steps and initiatives in various major interdisciplinary researches. Contrastingly, none of the Academies in Indian science have any economist fellow! The same picture is naturally reflected in the expert committees of the DST, CSIR, DAE, etc, who have failed to undertake any significant initiative in the emergent and important interdisciplinary research directions involving natural and social sciences. As mentioned already, this is clearly exemplified by the fact that the funding of the ISI does not come from any of the science funding agencies of the country. And all this even after seventy nine years of the pioneering, novel and successful experiment in this country at such a grand scale! In this connection, we may mention that Dr. T. Ramasami, present Secretary of the DST and ex-Director General of the CSIR, expressed (in a letter to one of us) keen interest in initiating the S.S. Bhatnagar Prize for Economic Sciences. Prof. S. K. Joshi, another ex-Director General of CSIR, too strongly feels (as expressed to one of us personally) the necessity of such an endeavor. Such a prestigious prize in economics from CSIR, if instituted, would clearly have profound impact not only in the Economic sciences but also in the promotion of interdisciplinary research involving natural and social sciences.

**Econophysics** is a new research field, which makes an attempt to bring economics in the fold of natural sciences or specifically attempts for a "physics of economics". The term Econophysics was formally born in Kolkata in 1995. The entry on Econophysics in The New Palgrave Dictionary of Economics, 2nd Ed., Vol 2, Macmillan, NY (2008), pp 729-732, begins with "... the term 'econophysics' was neologized in 1995 at the second Statphys-Kolkata conference in Kolkata (formerly Calcutta), India ...". The Econophysics research therefore formally completes **fifteen years of research** by the end of this year!

We believe, several important developments in Econophysics research have already taken place in the last one and half a decade. We now try to give a very brief (incomplete and biased!) list of some developments: (i) Empirical characterization, analyses and modeling of financial markets – in particular, the deviation from Gaussian statistics has been established, following the early observations of Mandelbrot and Fama (1960s) – beginning with the studies in 1990s by the groups of Stanley, Mantegna, Bouchaud, Farmer and others. (ii) Network models and characterization of market correlations among different stocks/sectors by the groups of Mantegna, Marsili, Kertesz, Kaski, Iori, Sinha and others. iii) Determination of the income or wealth distributions in societies, and the development of statistical physics models by the groups of Redner, Souma, Yakovenko, Chakrabarti, Chakraborti, Richmond, Patriarca, Toscani and others. The kinetic exchange models of markets have now been firmly established; this gained a stronger footing with the equivalence of the maximization principles of entropy (physics) and utility (economics) shown by the group of Chakrabarti. (iv) Development of behavioral models, and analyses of market bubbles and crashes by the groups of Bouchaud, Lux, Stauffer, Gallegati, Sornette, Kaizoji and others. (v) Learning in multi-agent game models and the development of Minority Game models by the groups of Zhang, Marsili, Savit, Kaski and others, and the optimal resource utilization "Kolkata Paise Restaurant" (KPR) model by the group of Chakrabarti. These have given important insights in such multi-agent collective parallel learning dynamics. In this context, it might be mentioned that to our knowledge, the KPR model might be the only model in Physics that is named after a city (namely Kolkata). Considerable literature has developed out of these and other studies [see the articles in this issue, and references therein]. We had invited all the above-mentioned groups and others to contribute their perspectives or views on these developments, and we are happy that most of them could. Unfortunately due to time constraints, several others could not contribute. The importance and proliferation of the interdisciplinary research of Econophysics is highlighted in this special issue of Science & Culture, which presents a collection of twenty nine papers (giving country wise perspectives, reviews of the recent developments and original research communications), written by more than forty renowned experts in physics, mathematics or economics, from all over the world.

We are glad that the entry on Econophysics in Encyclopedia of Complexity & System Science, Vol. 3, Springer, New York (2009) discusses some "influential" papers (p. 2803) from the "*Kolkata School*" (p. 2808; pp. 2800-2826), which exemplifies the role played by the Kolkata group of researchers in the development. Indeed it is gratifying to see that the revered physics journal Reviews of Modern Physics has published already their first review on Econophysics, namely on "Statistical mechanics of money, wealth, and income" by physicist Victor Yakovenko and economist J. Barkley Rosser, Jr. (Reviews of Modern

Physics, Vol. 81 (2009) pp. 1703-1725), which again highlights the contributions of the "*Kolkata School*". The unsatisfactory part of this marginal development in Econophysics or for that matter in the interdisciplinary researches in India has been the lack of any formal institutional support for such researches in emerging areas. Any such development intrinsically requires the participation of young researchers from both the disciplines. However, such topics are not a part of standard purview of either physics or economics and hence, no research fellowships or faculty positions are available in India yet. This is also due to the unfortunate lack of formal interactions between the communities of natural and social sciences.

It may be also mentioned that now formal and introductory courses in Econophysics are being offered by many distinguished universities like the ETH Zurich, the Casimir Research School, the Leiden University, and others. It is notable that the first faculty position in Econophysics has already been created in the Leiden University in Netherlands. Needless to mention the publications of several books, including the textbook *Econophysics: An Introduction* by S. Sinha, A. Chatterjee, A. Chakraborti and B.K. Chakrabarti (Wiley-VCH, Berlin, 2010) and monographs on Econophysics published by the distinguished presses like Cambridge University Press (e.g. *Introduction to Econophysics*, by R.N. Mantegna and H.E. Stanley, CUP, Cambridge, 1999), Oxford University Press (e.g., *Theory of Financial Risk and Derivative Pricing: From Statistical Physics to Risk Management*, by J.-P. Bouchaud and M. Potters, OUP, Oxford, 2003); Princeton University Press (e.g., *Why Stock Markets Crash: Critical Events in Complex Financial Systems*, by D. Sornette, PUP, Princeton, 2003), Springer, Elsevier, etc.

It is remarkable that at the end of 2009, George Soros funded the initiation of the Institute for New Economic Thinking as a

> "... think tank under the influence of the Financial crisis of 2007-2010. It has Nobel laureates George Akerlof, Sir James Mirrlees, A. Michael Spence and Joseph Stiglitz in its advisory board, as well as other top economists...." — Wikipedia (June, 2010)

With the backdrop of the present world economic scenario, India, with its mighty economic growth rate and position in Asia and the world, can hardly afford to remain indifferent to such global initiatives in the emerging interdisciplinary studies on the physical, computational and mathematical analysis and modeling of social dynamics and economics. This is all the more ironic in view of the pioneering initiatives and researches in the ISI structure! We now need even more vigorous efforts to create many such interdisciplinary research and teaching institutions, and university departments in the country to bring the Indian physical, mathematical and social scientific communities in a closer forum at the earliest; first to catch up with the initiatives already taken elsewhere and participate in the research and application developments at an equal footing, and then to put India in a leading role.

***Bikas K. Chakrabarti***
***Anirban Chakraborti***

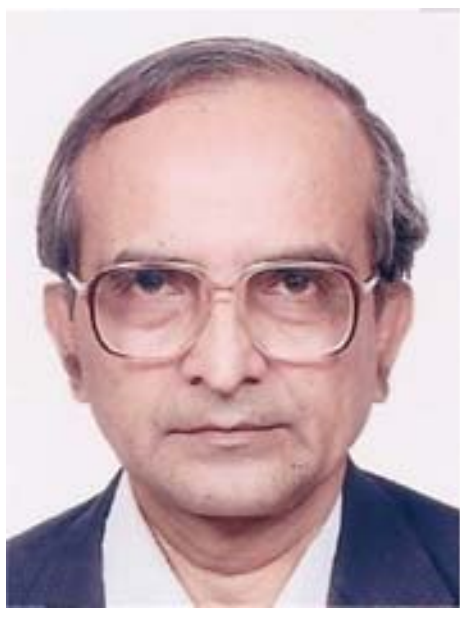

**Bikas K. Chakrabarti** is a senior professor of theoretical condensed matter physics at the Saha Institute of Nuclear Physics (SINP), Kolkata, and a visiting professor of economics at the Indian Statistical Institute, Kolkata, India. He received his doctorate in physics from Calcutta University in 1979 (for research at SINP). Following postdoctoral positions at Oxford University and Cologne University, he joined SINP in 1983. His main research interests include physics of fracture, quantum glasses, etc., and the interdisciplinary sciences of optimization, brain modelling, and econophysics. He has written several books (including publications by the Oxford University Press, Springer & Wiley) and reviews (including three in the Reviews of Modern Physics) on these topics. Professor Chakrabarti is a recipient of the S. S. Bhatnagar Award (1997), a Fellow of the Indian Academy of Sciences (Bangalore) and of the Indian National Science Academy (New Delhi). He has also received the Outstanding Referee Award of the American Physical Society (2010).

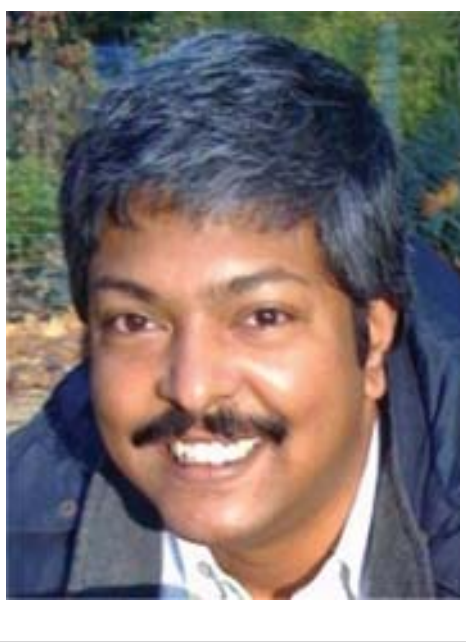

**Anirban Chakraborti** has been an assistant professor at the Quantitative Finance Group, Ecole Centrale Paris, France, since 2009. He received his doctorate in physics from Jadavpur University in 2003 (for research at SINP). Following postdoctoral positions at the Helsinki University of Technology, Brookhaven National Laboratory, and Saha Institute of Nuclear Physics, he joined the Banaras Hindu University as a lecturer in theoretical physics in 2005. Statistical physics of the travelling salesman problem, models of trading markets, stock market correlations, adaptive minority games and quantum entanglement are his major research interests. He is a recipient of the Young Scientist Medal of the Indian National Science Academy (2009).

**Editor's Note :** *This issue has been sponsored in part by the Centre for Appl. Math. & Comp. Sc., Saha Institute of Nuclear Physics, Kolkata*